\documentclass[12pt]{article}
\usepackage{epsfig}
\usepackage{amssymb}
\def\beq{\begin{eqnarray}}
\def\eeq{\end{eqnarray}}
\def\beqs{\begin{eqnarray*}}
\def\eeqs{\end{eqnarray*}}

\newcommand{{\SD}}{\rm SD}

\newcommand{\lan}{\langle}
\newcommand{\rrr}{\rangle}

\newcommand{\be}{\begin{equation}}
\newcommand{\ee}{\end{equation}}
\newcommand{\T}{\mbox{Tr}\> }

\def\centeron#1#2{{\setbox0=\hbox{#1}\setbox1=\hbox{#2}\ifdim
\wd1>\wd0\kern.5\wd1\kern-.5\wd0\fi
\copy0\kern-.5\wd0\kern-.5\wd1\copy1\ifdim\wd0>\wd1
\kern.5\wd0\kern-.5\wd1\fi}}
\def\ltap{\;\centeron{\raise.35ex\hbox{$<$}}{\lower.65ex\hbox{$\sim$}}\;}
\def\gtap{\;\centeron{\raise.35ex\hbox{$>$}}{\lower.65ex\hbox{$\sim$}}\;}
\def\gsim{\mathrel{\gtap}}

\begin{document}

\title{ \Large \bf Infinite statistics, symmetry breaking and combinatorial
hierarchy}

\author{V.Shevchenko \\
{\it  Institute of Theoretical and Experimental Physics},\\
{\it B.Cheremushkinskaya, 25 Moscow 117218 Russia} 
       }

\maketitle

\begin{abstract}

The physics of symmetry breaking in theories with strongly interacting quanta obeying infinite (quantum Boltzmann) statistics known as quons is discussed. The picture of Bose/Fermi particles as low energy excitations over nontrivial quon condensate is advocated. Using induced gravity arguments it is demonstrated that the Planck mass in such low energy effective theory can be factorially (in number of degrees of freedom) larger than its true ultraviolet cutoff. Thus, the assumption that statistics of relevant high energy excitations is neither Bose nor Fermi but infinite can remove the hierarchy problem without necessity to introduce any artificially large numbers. Quantum mechanical model illustrating this scenario is presented.

\end{abstract}

\section{Introduction}

A truly remarkable phenomenon in Nature is
coexistence of a few vastly different scales in one theoretical
framework. Sometimes it is known as a problem of "large numbers"
\cite{dirac}. The best known and physically the most interesting
example of this kind is given by the Standard Model. As is well
known this theory contains a tower of scales starting from
electron mass $m_e=511 \>\mbox{keV}/c^2$ and going up to QCD scale
$m_p = 938 \>\mbox{MeV}/c^2$, weak scale $m_W = 80 \>\mbox{GeV}/c^2$
and finally ending by the Planck scale $M_P=1.2 \cdot 10^{19} \>
\mbox{GeV}/c^2$ which represents, as many physicists believe, the
ultimate ultraviolet edge of our world.\footnote{We leave aside the
lower part of this tower, corresponding to smaller energies relevant
for condensed matter physics, chemistry and biology.} Despite each
large ratio in this tower calls for explanation, of particular
interest is the parameter $M_P$ in the weak scale units, which is
given by the number of order $10^{17}$. The danger this large
number is for the stability of radiative corrections to Higgs
boson mass has been widely discussed in the literature (see recent
review \cite{guidice}).

An interesting approach to the hierarchy problem known as TeV-scale gravity
and extra-dimensions scenarios came about a decade ago. These models
assume that the geometric properties of our familiar (3+1)
dimensional space-time change at some scale $L$, which is supposed
to be much larger than $L_P = 1/M_P$ and perhaps of (1-2 TeV)$^{-1}$
range. This change can be accompanied by appearance of some
additional particles, for example of the Kaluza-Klein type. Among
attractive features of these models is emergent nature of the Planck
scale $L_P$. In the original proposal \cite{add1,add2} the
fundamental ratio between $L_P$ and $L$ takes the form \be
\left(\frac{L}{L_P}\right)^2 \sim \frac{V_n}{L^n} \label{eq1} \ee
where $L$ is electroweak scale, $n$ - number of compact extra
dimensions and $V_n$ - their volume. Soft SM fields propagates only
in 4 dimensions, while gravitons feel the full $4+n$ dimensional
geometry.  In this scenario the smallness of the ratio $L/L_P$
follows from the large (in units of $L$, i.e. sub-millimeter for the
most phenomenologically interesting case $n=2$) size of extra
dimensions. The picture suggested in \cite{rs1,rs2} is different:
the model with one warped extra dimension generates the hierarchy
\be \log \frac{L}{L_P} \sim  \frac{\pi r_c}{L^{(5)}_P} \label{eq2}
\ee where $\pi r_c$ is the size of compact extra dimension and
$L^{(5)}_P$ - fundamental five dimensional Planck length. The
logarithmic function maps huge hierarchy between $L$ and $L_P$ into
much weaker hierarchy between $r_c$ and $L^{(5)}_P$. In other words
the weakness of gravity in this approach follows from the fact that
only exponential tail of the full graviton's wave function can be
seen by four-dimensional observer.

In a broader prospective, both these scenarios
reduce hierarchy of the mass scales to some geometrical hierarchy.
However, in functional terms, (\ref{eq1}) and (\ref{eq2}) are quite
different: while the power law (\ref{eq1}) makes $L_P$ small
introducing some large artificial scale $R$ - extra dimension size:
$L / L_P \sim (R/L)^{n/2}$; expression (\ref{eq2}) corresponds to
exponential increase: $L/L_P \sim \exp \> \gamma$, where huge number
$L/L_P$ is mapped into not so huge number $\gamma$. The physical
reason for appearance of the exponent in \cite{rs1} is quantum
tunneling of gravitons.

Recently the extra-dimensional approach to the hierarchy problem has
been given a new interpretation in \cite{gia2} (and further
developed in \cite{gia3,gia4}). It is based on the
following physical idea (also discussed in different contexts in
\cite{jakobson,veneziano,fursaev,huang,hsu}): suppose that above some scale
$L$ there is a dramatic rise of multiplicity, i.e. the number of
relevant degrees of freedom (d.o.f.) $N$ becomes huge at energies $E \gsim
1/L$. Then one can argue, both on perturbative and on
nonperturbative grounds, that the number of stable species with
typical mass $M$ must not exceed the following bound: \be NM^2
\lesssim M_P^2 \label{b} \ee up to some logarithmic corrections.
Another incarnation of the same idea takes a single particle
of mass $M$ carrying exactly conserved quantum number of
periodicity $N$, for example, $Z_N$ gauge symmetry charge. Then the
gravitational cutoff in such theories goes down to $M_P/\sqrt{N}$
and can be lowered to a TeV scale (thus solving the hierarchy
problem, or, at least, giving it a completely new prospective) if
one takes $N$ of the order $10^{32}$.

The bound (\ref{b}) can be given a natural interpretation from
Sakharov's induced gravity point of view (\cite{s}, see
\cite{adler,shapiro,vis} for introduction into the subject). Indeed,
typical contribution to Einstein-Hilbert gravitational Lagrangian
from one-loop effective action of matter particles in
curved space can be written as: \be M_P^2 = \frac{1}{G} =
\frac{1}{G_0} - \frac{1}{2\pi} \T_s \left[ M^2 - \mu^2 \log\left(
\frac{M^2}{\mu^2}\right) \right] \label{sak} \ee where $G_0$ is
"bare gravitational constant" (taken to be equal to infinity in
original Sakharov's approach), trace $\T_s$ is taken over the
spectrum with the corresponding numerical factors accounting for
particle content of the theory and $M$ is ultraviolet cutoff. If
there are $N$ particle species of a given type\footnote{Say, $N$
light noninteracting scalars. Of course, if both bosons and fermions
present in the theory, their contributions partly cancel each other.} in the
theory, the trace scales as $-\T_s[..]\propto N$. Then, making
original Sakharov's one-loop dominance assumption, one has from
(\ref{sak}): $1/G \sim NM^2$, i.e. just eq.(\ref{b}). The physical
interpretation of this result is clear. The rigidity of space, i.e.
its resistance against an attempt to curve it is proportional to the
number of particle species living in quantum vacuum in this space
because the curvature costs energy. Roughly speaking, more different
particle types the theory contains, weaker is the gravity in this
theory,\footnote{Again, leaving aside supersymmetric-like
cancelations.} i.e. rich and non-degenerate spectrum tends to wash
space-time distortions out.

Using alternative language for the same physics one can relate the induced contributions to
$M_P$ to spectral density function for the trace of energy-momentum tensor \cite{adler2}:
\be \frac{1}{G} =  i \>\frac{\pi}{6} \int d^4 x x^2 \lan 0| {\cal T} (T(x) T(0))|0 \rrr
\ee where $x^2 = x_0^2 - {\vec{x}}^2$ and $T(x)$ = $T^\mu_\mu(x) - \lan 0| T^\mu_\mu(x) |0 \rrr $.
If Green's function $\lan 0| {\cal T} (T(x) T(0))|0 \rrr$ scales at small distances, according to Wilson OPE, as  $\lan {\cal O}_0 \rrr / (x^2)^{4}$ (up to logarithms), the integral diverges quadratically and true
UV cutoff corresponds to $M_P \cdot \lan {\cal O}_0 \rrr^{-1/2}$. With extremely large $\lan {\cal O}_0 \rrr$, proportional to the number of species, it can be as small as a few TeV.

It is clear at the same time that introduction of the numbers like $10^{32}$ in the fundamental theory does not look quite natural. One therefore is interested in models where this rise is emergent rather than postulated.
The main goal of this paper is to discuss a scenario where this huge rise in number of relevant excitations appears in somewhat natural way. The key physical idea is a change of statistics relevant d.o.f. obey - from Bose/Fermi at low energies to infinite (quantum Boltzmann) one at high energies.
The paper is organized in the following way. In section 2 we remind the reader, using conventional theories as examples, how ultraviolet cutoff in low energy theory "knows" about the number of d.o.f. at high energies. In section 3 qualitative explanation of our scenario is given, while section 4 is devoted to simple quantum mechanical model, illustrating its main features. This section also present a summary of quon statistics and related issues.

\section{Phase transitions and multiplicity change}

We see that the change with energy in the structure of d.o.f. is a crucial point in the picture discussed above.
Therefore it is interesting to discuss theoretical schemes where this change is  built in theory structure dynamically. The simplest known field theoretical example is given by
Nambu $–$ Jona-Lasinio model. Fundamental Hamiltonian of this theory contains $N$ fermion d.o.f.: $H = H[{\psi}_i^\dagger , \psi_i]$, $i=1...N$. If chiral symmetry is broken, relevant low energy d.o.f. are pseudoscalars:
$H_{eff} = H[\Phi^\dagger, \Phi]$. The formal method to get $H_{eff}$ out of $H$ is well known bosonization
trick \`{a} la  Hubbard - Stratonovich: \be H[{\psi}_i^\dagger , \psi_i] \to H_0[\Phi] + H_0[{\psi}_i^\dagger , \psi_i] + H_{int}[\Phi,{\psi}_i^\dagger,\psi_i] \label{hs} \ee with subsequent integration over fermions. The phenomenon of the number of d.o.f. reduction at low energies is encoded in the term $H_{int} \sim \Phi \cdot \psi_i^\dagger \psi_i$.

The same logic applies to large $N_c$ QCD. Let us take $SU(N_c)$ Yang-Mills gauge theory with
two flavors of light fundamental quarks $u$ and $d$. There are $2N_c
$ spinor degrees of freedom in the theory with $2$ staying for the
number of flavors and $N_c$ for the number of colors. At low
energies the relevant excitations are three light pseudoscalar pions
$\pi^+$, $\pi^-$, $\pi^0$ and the corresponding lowest order
effective Lagrangian reads: \be L_{eff} = \frac14 F^2 \T
\left(\partial_\mu U^\dagger
\partial^\mu U\right) + \frac12 F^2 B \T \left( m \left(U + U^\dagger\right)\right)
\label{chiral}\ee where the $SU(2)$ matrix field $U$ is expressed in
terms of the pion fields $\vec \pi(x)$ $=$
$\left(\pi^1(x)\right.$, $\pi^2(x)$, $\left.\pi^3(x)\right)$, mass matrix
$m=\mbox{diag}(m_u, m_d)$, and $F$ is just pion decay constant (up
to ${\cal O}(m)$ corrections). The Lagrangian rewritten in terms of
the pion fields takes the following form (neglecting ${\cal
O}(\vec\pi^4)$ terms): \be L= (m_u + m_d) F^2 B + \frac12 (\partial
{\vec{\pi}})^2 - \frac12 (m_u + m_d ) B {\vec{\pi}}^2 \label{eff}\ee
with the obvious assignment $m_\pi^2 = (m_u + m_d)B $ and Gell-Mann - Oakes - Renner relation \cite{gor}: $ (m_u+m_d)\lan \bar{q}q\rrr = - F^2
m_{\pi}^2 $. It is worth noticing that despite the
Lagrangian (\ref{eff}) does not contain $N_c$ explicitly, the low
energy description as such is valid up to pion momenta smaller than
ultraviolet cutoff $\Lambda \sim 4\pi F$. The latter is
$N_c$-dependent quantity (one can formally remove
$N_c$-dependence from the Lagrangian (\ref{chiral}) by
redefinition of the pion fields but we prefer not to do it). Indeed,
taking into account that quark condensate scales linearly with $N_c$
and $m_{\pi}\to \mbox{const}$ in large $N_c$ limit, one has: $F^2 \sim N_c $. Thus induced contribution of pions to the
gravitational constant according to (\ref{sak}) scales
as\footnote{At large $N_c$ pure gluon contribution which scales as
$N_c^2$ starts to dominate, but it is not seen at low energies until
$\Lambda$ becomes of the order of the lowest glueball mass.} \be
\delta \left(\frac{1}{G}\right)_{pions} \sim \Lambda^2 \sim N_c
\label{sak2} \ee
despite there are only three and not $N_c$ pions in the theory. Notice also the change of statistics, common for the theories with nontrivial vacuum of this kind:
the fundamental d.o.f. are fermions, while low energy d.o.f. are bosons.
The meaning of (\ref{sak2}) is the same as has been
just discussed: the ultraviolet cutoff of the low energy theory
encodes information about the number of degrees of freedom in the
"fundamental" theory.\footnote{For $\Lambda$ larger than masses of
other, non-Goldstone hadrons, they have to be included as well.}
More degrees of freedom (associated with color in the considered
example) the underlying theory contains, weaker its low-energy
excitations interacts with the gravity. This is exactly original
idea behind \cite{gia2} and it is not surprising that (\ref{sak2})
is nothing than (\ref{b}) seen from a different prospective. Notice
that the "minimal charge" hypothesis discussed in \cite{motl} as a
consequence of "gravity is the weakest force" conjecture can be
rephrased as a statement that Planck mass $M_P$ in a theory with
dynamical scale $\lambda$ and coupling $g$ is never smaller than
$\lambda/g$. In large $N_c$ limit $g\sim 1/\sqrt{N_c}$ and we come
back directly to (\ref{b}) and (\ref{sak2}).

\section{Infinite statistics and quons}

The main problem with the mechanism of d.o.f. reduction at low energies discussed above is
its weak linear character. In other words, one gets as many d.o.f. as have been introduced into high energy theory from the beginning. Thus the gap between $n \sim$ a few dozens experimentally observed d.o.f.
at low energies and $10^{32}$ d.o.f. at TeV energies needed to solve the hierarchy problem seems rather artificial and in some sense replaces the mass hierarchy by even worse hierarchy with typical dimensionless parameter $10^{32}$.
It seems therefore reasonable to look at models with stronger than linear number of d.o.f. enhancement.
In the present paper, developing earlier results \cite{talk}, we discuss a quantum mechanical model where this goal is achieved by change of statistics relevant d.o.f. obey from conventional Bose/Fermi ones at low energies to the infinite (quantum Boltzmann) one at high energies. Thus the fundamental d.o.f. in our models are quons. The motivation for this choice and short summary of infinite statistics properties are presented below, but before we start systematic exposition it is useful to explain main qualitative aspects. A distinctive feature which makes quon theories different from conventional Bose/Fermi ones is the fact that typical quon Hamiltonian (even for free theory) contains interaction vertices of all orders. This is in contrast with the standard situation where correct choice of the vacuum and relevant excitations usually allows to  separate quadratic free part of Hamiltonian from high order terms, describing interactions. Thus having quon fields ${\cal A}_i^\dagger$, ${\cal A}_i^{\phantom{\dagger}}$
and Bose/Fermi effective low energy field ${\cal F}_j$ one could have in general instead of (\ref{hs})
\be
H_{int} \sim {\cal F}_j \cdot \biggl[ c_{11} {\cal A}_i^\dagger {\cal A}_i^{\phantom{\dagger}} + \sum_P c^P_{12} {\cal A}_i^\dagger {\cal A}_k^\dagger {\cal A}_k^{\phantom{\dagger}} {\cal A}_i^{\phantom{\dagger}} + ... \biggr] +  {\cal F}_j {\cal F}_l \cdot \biggl[ c_{21} {\cal A}_i^\dagger {\cal A}_i^{\phantom{\dagger}} +  ...  \biggr] + ...
\label{polik}
\ee
where $c^P_{\alpha\beta}$ are coefficients depending (for $\beta >1$) on particular permutation of the fields ${\cal  A}_i$ and the summation over permutations $\sum_P$ takes into account the fact that for quon fields ${\cal A}_i {\cal A}_k \neq \pm {\cal A}_k {\cal A}_i $. It is important that in general we do not expect to have any small parameter in the expansion (\ref{polik}), which could make lowest terms dominant. Therefore it is natural to expect that quon content of the true low energy ground state $|\Omega\rrr $ (which fixes the subset of dominant terms in (\ref{polik})) is determined dynamically by the high energy quon Hamiltonian $H[{\cal A}_i^\dagger, {\cal A}_i^{\phantom{\dagger}}]$. If one assumes that it is nontrivial superposition of $n$-quon states, one immediately gets stronger than exponential\footnote{And purely factorial in particular case of all $n$ quons being of different flavors.} in $n$ degeneracy for each low energy d.o.f. described by ${\cal F}_i$. Speaking differently, each low energy d.o.f. ${\cal F}_i$ interacts with (or, using more informal language, is made of) factorially large number of {\it different} quon d.o.f. represented by the products ${\cal A}_{i_1} {\cal A}_{i_2} ... {\cal A}_{i_n}$. These d.o.f. are condensed at low energy phase with ordinary fermions/bosons playing the r$\hat{\mbox{o}}$le of light excitations over this nontrivial vacuum, while they become relevant d.o.f. at high energies. In this way one can easily get exponential multiplicity, resulting in:
\be
\delta \left(\frac{1}{G}\right) \sim g_n M_n^2
\ee
where the factor $g_n$ of combinatorial origin scales with $n$ as $n!$ or even $n^n$ for totaly symmetric vacuum state.
Without intention to cook up numerical factors it is interesting to notice that numerically
one gets $M_{n} $ at TeV scale for $n=29\pm 1$. Since the number of all relevant d.o.f. in the SM at top quark mass scale is about one hundred, the "reduction" of hierarchy problem is even stronger than needed. It does not seem disappointing since the model is clearly too crude to pretend for quantitative predictions.
Nevertheless we find the proposed pattern attractive since the only "large" number one has to introduce in this case is the number of low energy "flavors", which is assumed to govern the low energy vacuum degree of degeneracy. In the next section we present a quantum-mechanical model where the features just discussed are explicit.

\section{Quantum mechanical model example}

To make our discussion self-contained, let us remind some basic facts about quon theories and infinite statistics.
The physical interest to theories with unconventional statistics is mainly motivated
by arguments coming from physics of black holes. Consider composite system built of $k$ free bosons with
masses $m_i$ placed into the volume $V$. In conventional theory this
system satisfies
Bose-Einstein statistics.\footnote{Of course, the same is true for
fermions and Fermi-Dirac statistics if $k$ is odd.} However this is
not the case if one includes gravity into consideration. Indeed, for
$k$ or $m_i$ or $1/V$ large enough in units of gravitational
constant $G$ a black hole forms. The latter is an object which satisfies neither
Bose-Einstein nor Fermi-Dirac statistics \cite{stro}. In quantum statistical
sense black
hole is identified by external observer as an object with
infinite number of internal states.  In other words, ideal gas of
black holes is a system,
whose many-body wave function is neither totally symmetric nor
antisymmetric with respect to the black holes permutation (another
example of this kind is given by gas of D0-branes \cite{minic}). Thus
one can say that heavy d.o.f. (i.e. the black holes)
satisfy infinite statistics and coexist with light d.o.f. (i.e. ordinary bosons or fermions) interacting with them in the course of absorption and emission.

 As is known \cite{dopli} there are only three types of
self-consistent statistics one is allowed to consider in four-dimensional field-theoretical
framework: parabosonic and parafermionic statistics, including
Bose-Einsteins and Fermi-Dirac ones, and quantum Boltzmann, or
infinite statistics, characterized by the relation \be a_i
a_j^\dagger = \delta_{ij} \label{q} \ee augmented by the Fock-state
representation defining relation $a_i|0\rrr =0$. Quantum fields
satisfying (\ref{q}) are conventionally called quons.
The
statistics (\ref{q}) was introduced in mathematical context in \cite{voc} and applied in noncommutative
probability theory, large-$N$ models, stochastic calculus etc (see, e.g. review \cite{ncpt}).
The first discussion of quon statistics in field
theoretical context is given in \cite{greenberg} and various aspects of quon theories have been
analyzed in many subsequent papers (see, e.g. \cite{govorkov,gree2}). In particular, physical interpretation of dark energy in infinite statistics language from holographic prospective is discussed in \cite{ng,jack,min2}.

 In quantum Boltzmann statistics $m$-particle state is constructed as \be |\phi^P_m \rrr = ( a_{i_1}^\dagger
)^{k_1} ( a_{i_2}^\dagger )^{k_2} ... ( a_{i_l}^\dagger )^{k_l}
|0\rrr \label{m} \ee with $k_1 + k_2 + ... + k_l=m$.  By index $P$ we denote the concrete permutation of the creation operators, which is necessary since different permutations correspond in general to different states. All states
have positive norm and can be normalized to unity by the condition
$\lan 0|0\rrr =1$. The states created by any permutations of
creation operators are orthogonal, i.e.
\begin{eqnarray} (a^\dagger_i .. a^\dagger_m |0\rrr )^\dagger \cdot
(a^\dagger_i ..
a^\dagger_m |0\rrr ) = \lan 0 | a_m .. a_i a^\dagger_i .. a^\dagger_m |0\rrr  = 1  \nonumber \\
 (a^\dagger_i .. a^\dagger_k .. a^\dagger_l .. a^\dagger_m |0\rrr
)^\dagger \cdot (a^\dagger_i .. a^\dagger_l .. a^\dagger_k ..
a^\dagger_m |0\rrr ) = 0 \;\; \mbox{for any } \; k\neq
l\end{eqnarray} For particles of the type $i$ one can define the
number operator $N_{i}$ such that \be N_{i} |\phi^P_m \rrr = k_i |\phi^P_m \rrr \;\; \mbox{and} \;\; [N_i, a_i]_{-} = - a_i \ee as \be N_i =
a_i^\dagger a_i + \sum\limits_l a_l^\dagger a_i^\dagger a_i a_l +
\sum\limits_{l,m} a_m^\dagger a_l^\dagger a_i^\dagger a_i a_l a_m +
... \label{n} \ee
or, in compact notation, $N_i = a_i^\dagger a_i + \sum_j a_j^\dagger N_i a_j$.
For free normal-ordered Hamiltonian one has $
H_0 = \sum\limits_{i=1}^n {\cal E}_0^i N_i $.
The condition (\ref{q}) automatically makes any product of quon operators normally ordered.
As is already pointed out, it is a distinctive property of quon statistics that free Hamiltonian is
given by an infinite series in creation and annihilation operators
(in contrast with conventional Bose or Fermi case where free
Hamiltonian is or can be made quadratic). Nevertheless there is no rich
dynamics in free quon theory due to (\ref{q}). Indeed, any state of
the kind (\ref{m}) or superposition of such states is an eigenstate
of the Hamiltonian with an eigenvalue $ \sum\limits_{i=1}^n
{\cal E}_0^{i} k_{i} $. This is nothing than the standard harmonic
oscillator equidistant spectrum.

It can be mentioning as a side remark that the structure of Hilbert space of states discussed in the present paper can be realized in alternative ways, i.e. without introducing quon d.o.f. Let us refer the reader to recent paper \cite{arzano}, where
the authors discuss momentum-dependent statistics resulted from $\kappa$-deformed Poincar\'{e} algebra.
In this approach the momentum splitting between, e.g. two-particle boson states $|p_1, p_2\rrr_\kappa $ and
$|p_2,p_1\rrr_\kappa$ is of the order of $|{\vec p}_1|\cdot |{\vec p}_2| / \kappa$ where $\kappa$ has the meaning of Planck mass $M_P$. This could be interpreted in terms of additional planckian internal degrees of freedom, which stay unresolved for energies much smaller than $M_P$. It would be interesting to reconcile those approach with the induced gravity paradigm.


In all theories based on quantum mechanics (and in the SM in particular)
any probability density must be symmetric
with respect to exchange of identical particles. While in nonrelativistic quantum mechanics this
symmetrization principle can be either postulated or derived under
some assumptions, in standard quantum field theory it directly follows from Bose or Fermi
commutation relations between field creation and annihilation
operators. In quon theories  there is no a priori any symmetry of
this type. For example, the states $ a_{1}^\dagger a_{2}^\dagger
|0\rrr $ and $ a_{2}^\dagger a_{1}^\dagger |0\rrr $ are orthogonal
and in free theory they are degenerate. It is reasonable to expect that in
interacting theory this degeneracy is lifted and the simplest way to get this is to associate interactions with permutations.
To be more concrete, we assume that symmetry breaking due to quon condensation takes place as described by the  following Hamiltonian (compare with (\ref{n})):
\be
 H = E (1-A_n^\dagger A_n + (A_n^\dagger )^2 (A_n)^2 + ... )
\label{deltah}
\ee
where $A_n$ is a superposition of linearly independent $n$-quon annihilation operators:
\be
A_n = ( \alpha_n^P \cdot a_{i_1} a_{i_2} ... a_{i_n} + \mbox{permutations} )
\label{aa0}
\ee
where for simplicity all indices are taken to be different. The energy scale $E$ is assumed to be large in the sense specified below. The coefficients $\alpha_n^P$ are chosen in such a way that the operators $A_n$ obey the same infinite statistics as $a_i$ does: $A_n A_n^\dagger = 1$. For totally symmetric combination it corresponds to permutation-independent $\alpha_n^P = 1/\sqrt{n!}$. It is worth noting that it is not necessary to take  $n$ in (\ref{deltah}) equal to the number of quon species $n$ or to omit terms with coinciding indices, since only general properties of $H$ discussed below are important, not particular choice of $A_n$.

With the choice (\ref{deltah}) the vacuum state at low energies is given by
\be|\Omega \rrr = A_n^\dagger |0\rrr \label{vac} \ee
with $H |\Omega \rrr = 0$. Notice that $|\Omega \rrr$ is not a vacuum for original "high-energy" quon d.o.f. in the sense that $a_i |\Omega \rrr \neq 0$. On the other hand it should be the vacuum for low energy d.o.f.: $f_i |\Omega \rrr = 0$.
The latter requirement can be reformulated if one  introduces instead of low-energy excitations $f_i$ over non-trivial vacuum  $|\Omega \rrr$ composite operators
$F_i^P = f_i \otimes a_{i_1}^\dagger a_{i_2}^\dagger ... a_{i_n}^\dagger$ acting on unbroken vacuum $|0\rrr$, where
the index $P$ reminds that different permutations of quon operators correspond, in general, to different
excitations. Then from (\ref{aa0}) and (\ref{vac}) one has
\be
\sum_P \alpha_n^P F_i^P |0\rrr = 0
\label{vv}
\ee
The crucial point now is concrete mechanism how (\ref{vv}) is realized.
Roughly speaking, in the effective theory language this depends on the way excitations
$f_i$ are made of excitations $a_i$. There are known explicit ways how to construct conventional fermion and boson operators out of quon ones \cite{gre3}, but what is important for us here is a structure of the sum (\ref{vv}).
We consider two extreme scenarios: 1) all terms but one are linearly independent and 2) $F_i^P |0\rrr = 0 $ for each transposition $P$. In the former case there is a unique non-degenerate vacuum in the theory and it is given by $|\Omega\rrr$. In the latter case each state $|\phi_n^P \rrr = a_{i_1}^\dagger ... a_{i_n}^\dagger |0\rrr$
(and any their linear combination) is annihilated by $f_i$: $f_i  |\phi_n^P \rrr = 0$. So the space of eigenstates of the Hamiltonian (\ref{deltah}) has the following structure in this case:
\be
|\Omega \rrr = \sum_P \alpha_n^P |\phi_n^P\rrr\;\; ;\;\;
|\omega^{P} \rrr = c^{P'} |\phi_n^{P'}\rrr + c^P |\phi_n^P\rrr
\ee
where $P'$ is some arbitrarily chosen but fixed transposition different from $P$ and
coefficients $c^{P}$ are chosen to have $\alpha_n^P c^P + \alpha_n^{P'} c^{P'} = 0$ and $\lan \omega^P | \omega^P \rrr = 1$. For permutation-invariant $\alpha_n^P = 1/\sqrt{n!}$ one has $c^{P'} = - c^{P} = 1/\sqrt{2}$. The vacuum has zero energy by construction: $\lan \Omega | H | \Omega \rrr = 0$ while
all states $|\omega^P \rrr$ are degenerate and separated from the vacuum by the gap $E$: $\lan \omega^P | H  | \omega^P \rrr = E$.

Thus we arrive to the following physical picture in this model. There is a ground state $| \Omega\rrr$
with $n$ low energy excitations obeying conventional statistics over it. The corresponding masses $m_i$ are assumed to be small at the scale of $E$. There are also $(n! - 1)$ states $|\omega^{P} \rrr$ with the excitations over them realized by the same $n$ different Bose/Fermi operators, which are all considerably heavier (since $E$ is assumed to be large). The high energy part of this tower is unseen in low energy processes characterized by the energies much smaller that $E$ with the only exception: it dominantly contributes to the low energy gravitational constant making it extremely small.

To summarize, one possible way to explain why gravity is so weak in typical particle physics units is to assume that
there is huge number of non-SM gravitationally interacting d.o.f., which are either quite heavy or do not interact with SM particles. It seems possible to explain this rise by change of statistics relevant degrees of freedom obey at high energies from Bose/Fermi to infinite one. In this scenario one gets typical scaling relation between true ultraviolet cutoff scale of low energy effective theory $M$ and inverse gravitational constant seen by low energy observer (which is Planck mass $M_P$) of the following form:
\be
M_P^2 \sim g_n M^2
\ee
where the factor  $g_n$ is of combinatorial origin and it scales with $n$ as $n!$ or even $n^n$ where $n$ is the number of low energy d.o.f. We presented simple quantum mechanical model illustrating this pattern, which we believe is typical for any quon theory with nontrivial vacuum.

\bigskip

{\bf Acknowledgments }

\medskip

The author acknowledges discussions with G.Dvali.
The work is supported by the INTAS-CERN
fellowship 06-1000014-6576 and partly by the grant for support of
scientific schools NS-843.2006.2; contract 02.445.11.7424 / 2006-112
and RFBR Grant 08-02-91008.

\end{document}